\documentstyle[pre,preprint,tighten,aps]{revtex}
\begin{document}
\draft
\title{Dynamic Multiscaling of the Reaction-Diffusion Front for $mA
+nB\to 0$.}
\author{Stephen Cornell,${}^{1,}${}\cite{currad}
Zbigniew Koza,${}^2$ and
Michel Droz${}^1$}
\address{
$^1$D\'epartement de Physique Th\'eorique,Universit\'e de Gen\`eve,
24 quai Ernest-Ansermet, CH--1211 Gen\`eve 4, Switzerland.}
\address{$^2$Institute of Theoretical Physics, University of Wroc\l aw,
pl.\ Maxa Borna 9, PL-50-204 Wroc\l aw, Poland.}
\date{Preprint cond-mat/9412044: 9th December 1994}
\maketitle
\begin{abstract}
We consider the reaction zone that grows between separated regions of
diffusing species $A$ and $B$ that react according to $mA+nB\to 0$, within the
framework of the mean-fieldlike reaction-diffusion equations.
For distances from the centre of the reaction zone much smaller than the
diffusion length $X_D\equiv \sqrt{Dt}$, the particle density profiles are
described by the scaling forms predicted by a
quasistatic approximation, whereas they have a diffusive cutoff at a distance
of order $X_D$.  This cutoff, and the power-law decay of the quasistatic
profiles, give rise to multiscaling behaviour, with anomalous values for the
exponents describing the moments of the density and reaction profiles.
Numerical solutions of the reaction-diffusion equations are in good
quantitative agreement with the predictions of this theory.
\end{abstract}

\section{Introduction}
The problem of a front that grows between initially separate regions
of diffusing species $A$ and $B$ that react according to
$mA+nB\to 0$ has provoked much recent interest
\cite{gara,jieb,chdr,koliko,codrch,codrch92,scstwi,arlahast,co}.
Studies have concentrated on the scaling properties of the reaction rate $R$
per unit volume
and particle density profiles $a$ and $b$, and on the critical
dimension above which the
mean-fieldlike reaction-diffusion rate equations are valid.
In geometries where the only spatial variation of the densities is along
the $x$-axis, these equations take the form\cite{gara,codrch}
\begin{eqnarray}
\partial_t a&=&D\partial_x^2 a - mka^mb^n\label{rateeq1},\\
\partial_t b&=&D\partial_x^2 b - nka^mb^n\label{rateeq2},
\end{eqnarray}
where $k$ is the reaction constant, and here and subsequently the reagents
are assumed to have the same
diffusion constant $D$.  The initial conditions appropriate to this problem
are
\begin{eqnarray}
a(x,0)&=&a_0\theta(x),\\
b(x,0)&=&b_0\theta(-x),
\end{eqnarray}
where $a_0$ and $b_0$ are constants, and $\theta$ is the Heavyside function.
The quantity $u\equiv
(a/m-b/n)$ obeys a simple
diffusion equation, with solution\cite{codrch}
\begin{equation}
u={1\over 2}\left({a_0\over m}-{b_0\over n}\right)-
{1\over 2}\left({a_0\over m}+{b_0\over n}\right){\rm erf}\left({x\over
2\sqrt{Dt}}\right).\label{usol}
\end{equation}
The reaction is concentrated in the region where the densities of the two
species are comparable.  The centre $x_f$ of the reaction zone
may be defined as the point
where $u=0$.  We therefore have
\begin{equation}
x_f=2({Dt})^{1/2}{\rm erf}^{-1}
\left[{\left({a_0/m}-{b_0/n}\right)\over\left({a_0/m} +
{b_0/n}\right)}\right],\label{a0mb0n}
\end{equation}
so if $a_0/m=b_0/n$ we have $x_f=0$
(n.b. in reference\cite{codrch}, $x_f$ was defined
as the point of maximal reaction, which is not necessarily
the same point).

If one assumes that the penetration of one species into the other
is much shallower than the diffusion length $(Dt)^{1/ 2}$,
the
reaction between the two species takes place within a distance
$ w\ll(Dt)^{1/2}$ of $x_f$.
One expects that, for $x\ll (Dt)^{1/2}$, the profiles will be described
by the single lengthscale $w$, leading to the following scaling hypothesis
\cite{gara,codrch}
\begin{eqnarray}
a(x,t)&=&t^{-\gamma}A\left({x-x_f\over t^{\alpha}}\right),\label{dyscale1}\\
b(x,t)&=&t^{-\gamma}B\left({x-x_f\over t^{\alpha}}\right),\\
R(x,t)&=&t^{-\beta}\phi\left({x-x_f\over t^{\alpha}}\right)
\label{dyscale3},
\end{eqnarray}
where $w\sim t^\alpha$, $\phi=A^mB^n$, and $\beta=(m+n)\gamma$.  The number
of particles of either species arriving at the reaction front is $\propto t^{-
{1/2}}$, which must equal the total reaction rate, so we must have
$\beta-\alpha={1/2}$.  Consistency of this scaling ansatz
with the equations of motion leads to
$\alpha={(1/2)}{(m+n-1)/(m+n+1)}$\cite{codrch}.

The related case of a  front formed by
opposing constant diffusion currents $J_A=mJ$ and $J_B=-nJ$
of $A$- and $B$-particles imposed at $x=-\infty$ and $+\infty$ respectively
has recently been studied\cite{bere,codr,leca}.
In this case, the system approaches a steady state where the
equations
\begin{equation}
(D/m)\partial_x^2 a=ka^mb^n=(D/n)\partial_x^2 b,\label{sseom}
\end{equation} and boundary
conditions may be written in dimensionless form, so that the
following scaling ansatz
is valid for all $x$\cite{codr}:
\begin{eqnarray}
R(x)&=&{J\over w}\phi_{ss}\left({x\over w}\right),\\
a(x)&=&J w A_{ss}\left({x\over w}\right)\label{sssa},\\
b(x)&=&J w  B_{ss}\left({x\over w}\right)\label{sssb},
\end{eqnarray}
where $w(J,D,k)\propto J^{-\nu}$ and $\nu={(m+n-1)/( m+n+1)}$.
{}From (\ref{sseom}), we have $\partial_x^2u=0$ [$u\equiv(a/m-b/n)$],
whose solution with these boundary conditions is $u=-Jx/D$.
For $(x/w)\gg 1$, the $B$-particles are overwhelmingly in the majority,
so one has $b=nJx/D+na/m\approx nJx/D$, and hence
$  A_{ss}^{\prime\prime}(y)\sim
( A_{ss})^my^n$, leading to
\begin{equation}
A_{ss}(y)\sim\cases { y^{-{n/4}}
\exp(-\sigma y^{1+(n/2)})&for $m=1$,\cr
y^{ -{(n+2)}/{(m-1)} }&for $m> 1$,\cr}
\label{aprof}
\end{equation}
as $y\to \infty$,
where $\sigma$ is a constant.
Similar results hold for $B_{ss}$ by interchanging $m$ and $n$.
Within this approach, it is also possible to show\cite{codr}
 that the `mean-field'
assumption $R=ka^mb^n$ (assumed in all the above equations) is
valid for microscopic stochastic systems  in spatial
dimension $d>d_c\equiv 2/(m+n-1)$.

For the time-dependent problem, when $x\gg w$
one of the species is overwhelmingly in the majority, so $|u|\to
{\rm max}(a/m,b/n)$, and the profile of the majority particle density is $\sim
|x-x_f|/t^{1/2}$ for $(Dt)^{1/2}\gg x\gg w$.
The diffusion current of particles arriving at $x_f$ is therefore
$J\sim t^{-{1/2}}$, and the characteristic timescale on which this
current changes is $(d\log J/dt)^{-1}\propto t$.  The equilibration time
of the front is of order $Dw^2$, so since $\alpha<{1/2}$ one would
expect that the reaction zone has enough time to reach the steady-state profile
it would have if the current $J$ were constant.  One would therefore predict
that the results of the steady-state problem, and hence the dynamic
scaling ansatz, would be applicable to the time dependent case for
$x\ll (Dt)^{1/2}$\cite{codr}.

For $m=n=1$ the scaling forms (\ref{dyscale1}--\ref{dyscale3})
have been proved rigorously to describe the
asymptotic behaviour as $t\to \infty$ of the reaction-diffusion
equations (\ref{rateeq1},\ref{rateeq2})\cite{scstwi}.
Experiments on real systems, and simulations of microscopic
stochastic models, also appear to verify the scaling theory and exponents
in dimension $d\ge 2$\cite{jieb,chdr,koliko}.
For $d=1$,
there has been some controversy as to whether the scaling theory is
valid\cite{codrch,arlahast},
but the most recent results\cite{co}
appear to show that the steady-state results
do indeed apply.  However, for
$(m,n)\ne (1,1)$, where rigorous mathematical results are
not available, the case is much less clear.
Numerical simulations of microscopic stochastic models
in $d=1$ are consistent with a scaling ansatz\cite{codrch92},
but they are of low
precision, since reaction events are much rarer than for $m=n=1$.
The fact that at least one of the particle density profiles must
decay algebraically (see Eq.\ (\ref{aprof}))
might invalidate the assumption that reactions take place
within a zone of width $w\ll (Dt)^{1/2}$.

In this paper, we shall give careful arguments to show that the scaling ansatz
is indeed valid for lengthscales much smaller that $\sim t^{1/2}$.
We shall then show that the two lengthscales in the problem,
$w$ and $(Dt)^{1/2}$, together with the
power law tails of the steady-state profile, give rise to a multiscaling form
for the particle density profile, whose moments are described by a spectrum
of exponents between $\alpha$ and ${1/2}$.
We then present numerical solutions of the reaction-diffusion equations,
and show that the results are in good agreement with the theoretical
predictions.

\section{Validity of the Scaling Ansatz}

The scaling ansatz can be shown to be exact for the case of a steady-state
front formed between balancing opposing currents, and so its applicability to
the time-dependent case relies on the front being formed quasi-statically
\cite{codr}.
Naively, one would expect that the time for a diffusive system to
equilibrate within a region of size $\sim t^{\nu}$ would be
$\propto t^{2\nu}$, whereas the timescale
upon which the current $J\propto t^{-{1/2}}$ changes is
$(d\log J/dt)^{-1}\propto t$, which predicts that
quasistatic approximation would be valid for lengthscales with
$\nu<{1/2}$.
However, since some of the density profiles decay algebraically,
one might wonder whether the flow of particles towards $|x|\to\infty$
necessary to sustain these steady-state profiles might be too great for
the quasi-static approximation to be valid.

In this section, we shall show that the quasistatic approximation is internally
consistent for lengthscales smaller than $t^{1/2}$,
in that it predicts that: (i) the number of particles up to a
distance $\sim t^{1/2}$ is always much less than the total particles that
have reacted, so that the number of particles in the tails is never too much to
have a feedback effect on the profiles at distances of order $t^{1/2}$;
(ii) the time taken for each part of the particle density tail to equilibrate
at its quasistatic value is always much less that the characteristic timescale
on which this value changes.  We shall then discuss what this implies about the
behaviour of the moments of the density and reaction profiles.

Consider the part of the
tail of the $A$-particle profile $a(x,t)\sim t^{-\gamma}
(x/t^\alpha)^{-\lambda}$, where
$\lambda=[n+2]/[m-1]$ (see Eq.\ (\ref{aprof})), in the region
$x_1<x<x_2$, with $x_1\propto t^{\epsilon_1}$ and $x_2\propto t^{\epsilon_2}$
($\alpha<\epsilon_1<\epsilon_2<{1/2}$).
The current of $A$-particles at $x$ is
\begin{equation}
J_A=-D\partial_x a
\sim {a(x,t)\over x},
\end{equation}
so the ratio $J_A(x_2)/J_A(x_1)=t^{-(\epsilon_2-\epsilon_1)(1+\lambda)}\to 0
$ as $t\to\infty$.  Almost all of the particles that enter at
$x_1$ are therefore removed by the reaction, rather than by diffusing out at
$x_2$.  The number of particles in the region is
\begin{equation}
N_A\equiv\int_{x_1}^{x_2}a\,dx
\sim\cases{a(x_1)x_1& for $\lambda>1$,\cr
           t^{\alpha-\gamma}\log\left(
          {x_2\over x_1}\right)& for $\lambda=1$,\cr
           a(x_2)x_2& for $\lambda<1$.}
\end{equation}
The number of particles in the tail can diverge for certain values
of $\lambda$.  This could invalidate the assumption that the total reaction
rate equals the number of particles arriving at the origin if this number were
found to be larger than the total number of particles $\propto t^{1/2}$
that have reacted.
However, the total number of particles in the tail up to a lengthscale
$t^{1/2}$, found by substituting $\epsilon_2={1/2}$, is
found in each of the above cases to be of order less than $t^{1/2}$.
The time taken for $N_A$
particles to enter the region $x_1<x<x_2$ is
\begin{equation}
N_A/J_A(x_1)\sim\cases{ t^{2\epsilon_1}& for $\lambda>1$,\cr
                        t^{2\epsilon_1}\log t& for $\lambda=1$,\cr
  t^{\epsilon_1+\epsilon_2 - (\epsilon_2-\epsilon_1)(1-\lambda)}&
    for $\lambda>1$,}
\end{equation}
which is always $\ll t$ since $\epsilon_1<\epsilon_2<{1/2}$.
The front therefore has enough time to reach its steady-state value for
lengthscales smaller than $t^{1/2}$.

The scaling ansatz (\ref{dyscale1}--\ref{dyscale3})
would therefore appear to be consistent for all lengthscales
$\sim t^\epsilon$, with $\epsilon<{1/2}$.
For $\epsilon\ge {1/2}$, the density of particles is limited by
diffusion, and so we expect there to be some kind of exponential
cutoff in all of the profiles on such lengthscales.  We therefore propose the
following ansatz for $a$ and $b$ in the limit $t\to \infty$:
\begin{eqnarray}
a(x,t)&=&a_{ss}(x,t)G_A
\left({x\over t^{1/ 2}}\right)\label{multiscale1},\\
b(x,t)&=&b_{ss}(x,t)G_B
\left({x\over t^{1/ 2}}\right),\label{multiscale2}
\end{eqnarray}
where $a_{ss}=t^{-\gamma}A_{ss}(x/t^\alpha)$ and $b_{ss}=t^{-\gamma}
B_{ss}(x/t^\alpha)$ are the solutions to the steady-state equations
(\ref{sssa},\ref{sssb}),
and $G_A(y)$ and $G_B(-y)$ are functions that provide a cutoff
at $y={\cal O}(1)$, and ensure that $a(x,t)$ and $b(x,t)$
satisfy (\ref{usol}) away from the reaction zone.
The actual form of $G_A(y)$ and $G_B(-y)$
is unimportant, provided that all moments of the tail for $y>0$
are defined and that there is no power-law behaviour for $y\to 0$.

This form leads to multiscaling behaviour for the moments of
the particle profiles, by virtue of the power law tails of $a$ and/or $b$
when $(m,n)\ne(1,1)$.
Consider a function $F$ of the form
$F(x,t)=t^{-\delta}\phi(x/t^\alpha)G({x/t^{1/2}})$, where $\phi(y)\to
y^{-\mu}$ as $y\to\infty$, $\phi\to 1$ as $y\to 0$, $G(y)\to 1$ as
$y\to 0$, and all positive moments of $G$ are defined. Then the
$q$'th moment of $F$ is of the form
\begin{eqnarray}
\int_0^{\infty} x^q F(x,t)\,dx
&=&\int_0^{\infty} x^q t^{-\delta}\phi\left(x\over
t^{\alpha}\right)G\left({x\over t^{1/2}}\right)\,dx\nonumber\\
&\sim& \cases {
t^{\alpha (q+1)-\delta}
 & for $\mu>q+1$,\cr
t^{\alpha (q+1)-\delta}\log t&for $\mu=q+1$,\cr
t^{\alpha\mu-\delta +{1\over 2}(q-\mu+1)}
& for $\mu<q+1$.
}\nonumber
\end{eqnarray}
When $\mu>q+1$, the $q$'th moment of $\phi$ is finite, whereas, for $
\mu<q+1$, the dominant contribution comes from $(x/t^{1/2})={\cal O}(1)$.
Defining $X^{(q)}\equiv [\int x^qF\,dx/\int F\,dx]^{1/q}$, we find that, for
$\mu\le 1$, $X^{(q)}\sim t^{1/2}$ (with logarithmic corrections for
$\mu=1$).  For $\mu<1$, we find the multiscaling behaviour
$X^{(q)}\sim t^{\zeta_q}$, with $\zeta_q$ increasing monotonically as a
function of $q$ from $\alpha$ to $1/2$.

By substituting from (\ref{aprof})
the appropriate power-law tails of $A_{ss}$ and $B_{ss}$, the multiscaling
forms predict the following behaviour for the following quantities
(without loss of generality, we have assumed $m\ge n$):

\begin{eqnarray}
w^2&\equiv& { \int_{-\infty}^{\infty} x^2 R(x,t)dx\over
\int_{-\infty}^{\infty} R(x,t)dx}\nonumber\\
&\sim&\cases{
t^{2\alpha}& for $m-n<3$,\cr
t^{2\alpha}\log t& for $m-n=3$,\cr
t^{1-\left({1\over 2}-\alpha\right)(\nu-1)} & for $m-n>3$,}\label{w2}\\
\noalign{\smallskip}
w_a^2&\equiv&{\int_{-\infty}^{x_f} x^2 a(x,t)dx\over
\int_{-\infty}^{x_f}  a(x,t)dx}\nonumber\\
&\sim&\cases{
t^{2\alpha}& for $3m-n<5$,\cr
t^{2\alpha}\log t & for $3m-n=5$,\cr
t^{1-\left({1\over 2}-\alpha\right)(\lambda-1)}& for $3m>n+5>m+2$,\cr
{t\over\log t}& for $m-n=3$,\cr
t& for $m-n>3,$
}
\\
\noalign{\smallskip}
w_b^2&\equiv&{\int_{x_f}^{\infty} x^2 b(x,t)dx\over
\int_{x_f}^{\infty}  b(x,t)dx}\nonumber\\
&\sim&\cases{
t^{2\alpha}& for $m+5>3n$,\cr
t^{2\alpha}\log t & for $m+5=3n$,\cr
t^{1-\left({1\over 2}-\alpha\right)(\kappa-1)}& for $m+5<3n$,
}\label{wb2}
\end{eqnarray}
where  $\lambda=(n+2)/(m-1)$, $\nu=2+\lambda$, and $\kappa=(m+2)/(n-1)$.
If these quantities were described by a one-length scaling theory,
all of these quantities would behave as $\sim t^{2\alpha}$, so we
describe departure from this behaviour as `anomalous'.

Defining
\begin{equation}
x_c\equiv {\int_{-\infty}^{\infty} x R(x,t)dx\over
\int_{-\infty}^{\infty} R(x,t)dx},\label{xc}
\end{equation}
a similar procedure may be used to find the scaling behaviour for
$x_c$.  Notice, however, that the contribution to $x_c$ coming from the
scaling term is identically zero, since $mR_{ss}=D\partial_x^2 a$ implies
$\int_{-\infty}^\infty xR_{ss}\,dx=0$.  The behaviour of $x_c$ is therefore
determined wholly by the corrections to scaling.
{}From (\ref{rateeq1},\ref{rateeq2}), one has
\begin{eqnarray}
\int_{-\infty}^\infty xR\,dx&=&\int_{-\infty}^0 {D\over m}x\partial_x^2 a\,dx
+\int_0^{\infty} {D\over n}x\partial_x^2 b\,dx\nonumber\\
&& -
\int_{-\infty}^0 x\partial_t a\,dx
-\int_0^{\infty} x\partial_t b\,dx.
\end{eqnarray}
The first two terms may be integrated by parts, yielding $D[a(0,t)/m-
b(0,t)/n]$, which is zero by virtue of (\ref{usol})
for the initial conditions
$u(0,0)=0$.
The final two terms are of opposite sign, and cancel identically for $m=n$
by symmetry.  For $m\ne n$, they typically have different scaling behaviours,
and so the scaling behaviour is determined by the largest term.
Substituting the forms (\ref{multiscale1},\ref{multiscale2})
for $a$ and $b$, and differentiating, one finds the following
scaling behaviour for $x_c$.
\begin{equation}
x_c\sim\cases{
t^{3\alpha - 1}& for $ 2m-n<4$,\cr
t^{{1\over 2}-{1\over m-1}}\log t & for $ 2m-n=4$,\cr
t^{\alpha-\left({1\over 2}-\alpha\right)\lambda}& for $2m-n>4.$
}
\end{equation}

\section{Numerical Simulations}

In order to verify the scaling ansatz (\ref{dyscale1}--\ref{dyscale3})
we have solved
the reaction-diffusion equations
(\ref{rateeq1},\ref{rateeq2}) numerically.
We approximated (\ref{rateeq1}) and (\ref{rateeq2})
with a finite-difference method, with
$\Delta t$, $\Delta x$ satisfying $k \Delta t = 0.01$, $D \Delta t /
(\Delta x)^{2} = 0.04$, and a lattice of $N = 12001$ sites, which was
sufficient for finite-size effects to be unimportant. Henceforth we
will
choose the reference system in which $\Delta x = \Delta t = 1$, so
that
our numerical results will correspond to the case $k = 0.01$ and
$D=0.04$. The initial condition satisfied
$a_0/m =b_0/n$, so that $x_f=0$ from (\ref{a0mb0n}).
The reaction rate $R$ thus
obtained for times $t = 10^3\ldots10^7$ and $n = 1$, $2 \le m \le 4$
are
presented in Figures 1--3, rescaled by $t^{\alpha+{1\over 2}}$ and plotted
as a function of $x/t^\alpha$.  The solid curve in each figure is a numerical
solution the steady-state equation (\ref{sseom}), for the same values of
$D$ and
$k$, and using the time dependent current as boundary condition.
We see that the data appear to converge to the solid line as $t\to\infty$,
so that the scaling ansatz (\ref{dyscale1}--\ref{dyscale3}) is valid in the
sense that
\begin{equation}
t^{\alpha+{1\over2}}R(x/t^\alpha)\to F(x/t^\alpha) \qquad\hbox{as }t\to\infty,
\label{scaleass}
\end{equation}
where $F(x/t^\alpha)$ is a function of $x/t^\alpha$.
Notice that the convergence appears to be slower for larger values of $m$.

Another interesting feature of Figures 1--3 is that the
point at which $R$ reaches its maximal value differs from 0. As
the $x$-axes of these figures have been rescaled by a factor $t^\alpha$, the
location
of this point changes with time like $t^\alpha$.  This shows that the
definition of $x_f$ as the point of maximal reaction is not equivalent to
the definition in this article as the point where $u=0$.

To verify the multiscaling properties of the reaction zone, we
investigated
the behaviour of the functions $w$, $w_a$, $w_b$, and $x_c$. According to
(\ref{w2}--\ref{xc}),
they should diverge as $t^\nu \log^\kappa(t)$, with $\nu$
and
$\kappa$ being some exponents dependent on $m$ and $n$.
In Figs.\ 4--7 we have plotted these quantities, rescaled by
$\log^{-\kappa}t$,
on a log-log scale, using the theoretical values of $\kappa$.
The values of $\nu_w$, $\nu_a$, $\nu_b$, and $\nu_c$
(corresponding to the
behaviour of the properties $w$, $w_a$, $w_b$, and $x_c$ respectively)
were estimated
from a least-squares fit to the last decade in these figures, and are compared
with the theoretical values in Table 1.  Agreement for most of the exponents
is very good.  Since $n=1$, the theory predicts that $\nu_b$ assumes its
non-anomalous value $\alpha$, and the
measured values are in close agreement.  The values for $\nu_a$ are
anomalous, and are also well reproduced.
The values for $\nu_w$ and $\nu_c$ are close to the
theoretical predictions for $m=2$, but deviate for $m=3$ and 4.

The reason why some of the exponents deviate from the theoretical values
may be understood from Figs.\
1--3.  The convergence to the solid curve for $x<0$, where Eq.\ (\ref{aprof})
predicts an exponential decay, is very rapid---semilog plots of $R(-x)$ were
found to be in very good agreement with these predictions for up to 10 decades.
However, the convergence of the profiles to the steady-state profiles is much
slower for $x>0$ (where the asymptotic behaviour is algebraic).
In fact, numerical investigation of the steady-state
profiles showed that the regime for which the power law behavior predicted
by (\ref{aprof}) appears is beyond the point at which the diffusive cutoff
is already active in the data.  This means that the asymptotic regime
has not yet been reached.  In view of this fact, the agreement between the
measured exponents and the theory is surprisingly good.

\section{Conclusions}

The multiscaling theory predicts that the reaction profile of the
system is described asymptotically by
a scaling form, in the sense that Eq.\ (\ref{scaleass}) holds,
but that the moments of the reaction and density profiles
may have anomalous behavior.  Numerical solutions of the reaction-diffusion
equations verify the asymptotic scaling behavior, and also give values for
the anomalous exponents close to those predicted by the theory.  Longer times
would have to be simulated to find better values for the exponents.

The convergence to the asymptotic behavior becomes progressively slower
as the order of the reaction is increased.  This means that simulations
probing the asymptotic behavior also become more difficult.  Nevertheless,
Eq.\ (\ref{w2}) suggests that it would be worthwhile to look at at least
one  case
where $m-n>3$.

In the steady-state problem, it has been shown\cite{codr} that the critical
dimension is $d_c=2/(m+n-1)$.  This means that the reaction-diffusion equations
correctly describe the scaling behavior of `real' stochastic realizations
for all physical dimensions, except for $(m,n)=(2,1)$ in dimension 1, where
logarithmic corrections to the steady-state behavior are expected.
Because  of the strong link between the steady-state problem and the
time-dependent problem studied in the present article, we expect the
critical dimensions to be the same.
The only microscopic simulation results available \cite{codrch92} agree
broadly with the picture in the present article, though they are not of
sufficiently high quality to give a thorough test.

The theory outlined in this paper is based on heuristic arguments and
numerical results only, so a first-principles analytical justification is
needed. It would be possible to write the equation of motion for the
corrections to the multiscaling terms, and then investigate whether they are
truly small and do not contribute to the behavior of the moments.
Although no conclusive results have been found, a preliminary attempt at this
procedure suggests that there may be values for $m$ and $n$ where the
assumption may break down that the interpenetration of the species is
smaller than the number of particles that have reacted\cite{wit}.  The
multiscaling theory presented in this paper might therefore have to be
revised in those cases.

\section*{Acknowledgements}

ZK was supported by KBN Grant no.\ 2P 302 181 07.
MD acknowledges support from the Swiss National Science Foundation.
We would like to thank Peter Wittwer for stimulating discussions.

\begin{table}[h]
\caption{
The measured values of the exponents $\nu_w$, $\nu_a$ and $\nu_b$ describing
$w$,  $w_a$, $w_b$ and $x_c$ respectively, together with
the non-anomalous exponent $\alpha$, for three values of the
duple $(m,n)$.  The values in square brackets are
the predictions of the multiscaling theory; the presence of a value for
$\kappa$ indicates the presence of a logarithmic correction of the form
$\log^\kappa(t)$, account of which has been taken in calculating the $\nu$.
}
\begin{center}
\begin{tabular}{clll}
 Quantity    & $\phantom{-}$(2,1)   &      (3,1)   &   (4,1)   \\ \hline
$ \alpha$ &$ \phantom{-}{1\over 4}$ & ${3\over10}$ & ${1\over3}$    \\
$  \nu_w$  & $\phantom{-}0.26\quad [{1\over 4}]$ & $ {0.35}\quad[{3\over 10}]
$& $0.37 \quad[{1\over 3}, \kappa={1\over 2}]$     \\
$ \nu_a $ &$\phantom{-}{0.27} \quad[{1\over 4}, \kappa={1\over 2}]$      &
$  {0.44}\quad [ {0.45}]$    & $0.50\quad [{1\over 2}, \kappa={-{1\over 2}}]
$  \\
$  \nu_b  $&$\phantom{-}{0.25} \quad[{1\over 4}]$      &$
{0.29}\quad[{3\over 10}]
$   & ${0.32}\quad [{1\over 3}] $  \\
$  \nu_c $  & ${-0.25}\quad[{-{1\over 4}} ]$    &
$0.06\quad[0] $ &$ {0.21}\quad[{1\over 6}]$
\end{tabular}
\end{center}
\end{table}

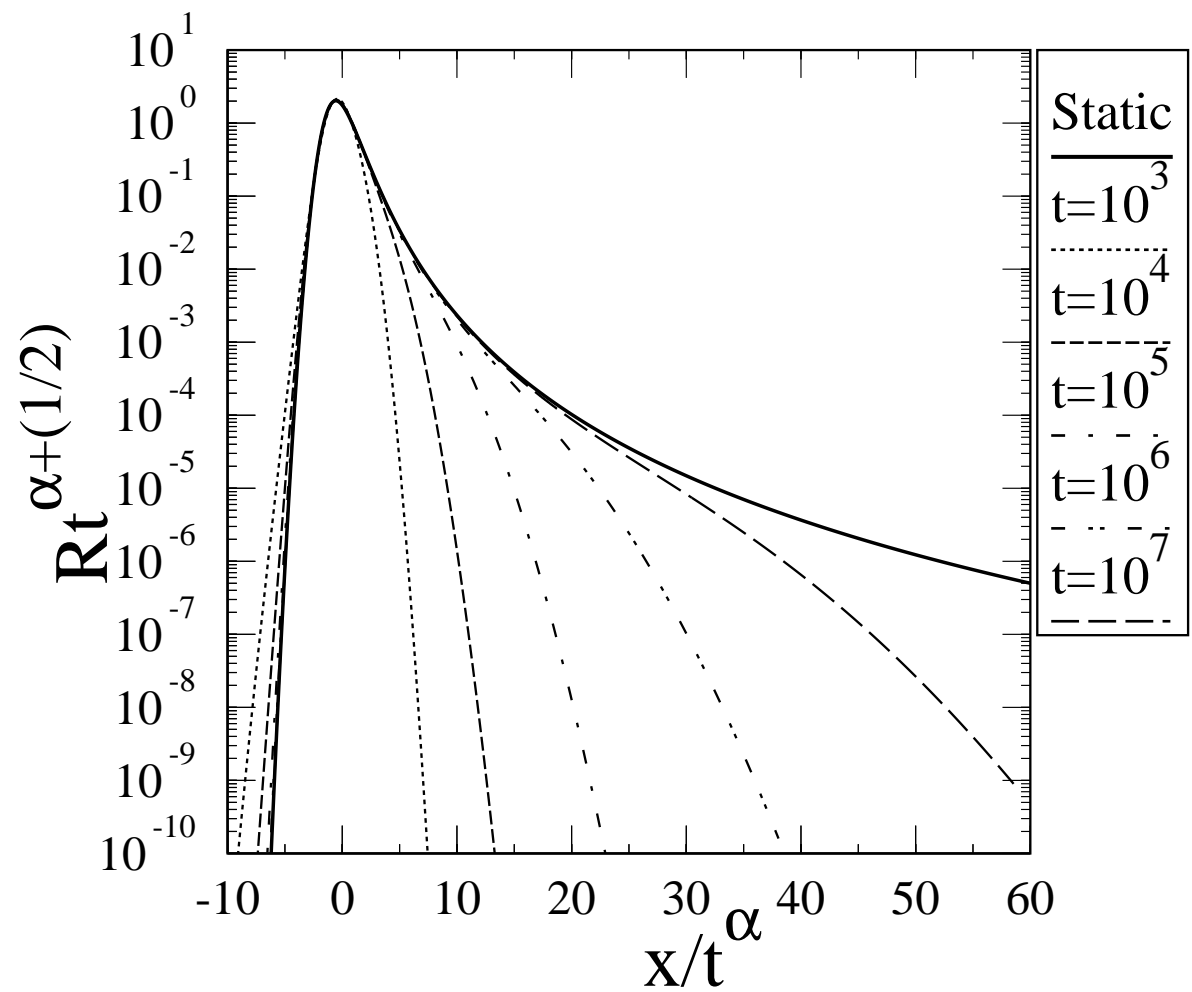
\begin{figure}
\caption{$R \cdot t^\beta$ vs.\ $x/t^\alpha$ for $n = 1$ and $m =2$
and $t = 10^3, 10^4,\ldots,10^7$.}
\end{figure}

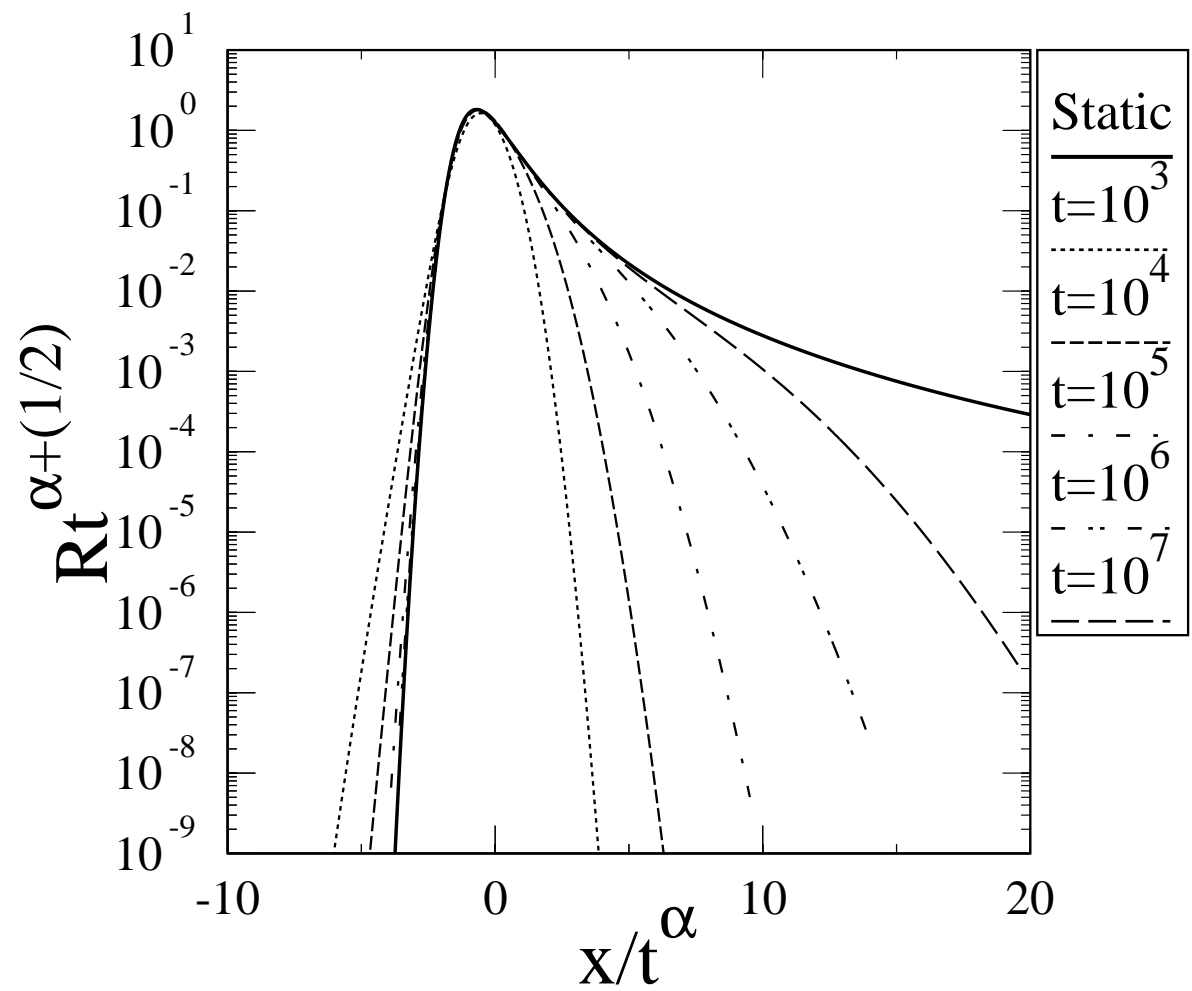
\begin{figure}
\caption{$R \cdot t^\beta$ vs.\ $x/t^\alpha$ for $n = 1$ and $m=3$
and $t = 10^3, 10^4,\ldots,10^7$.}
\end{figure}

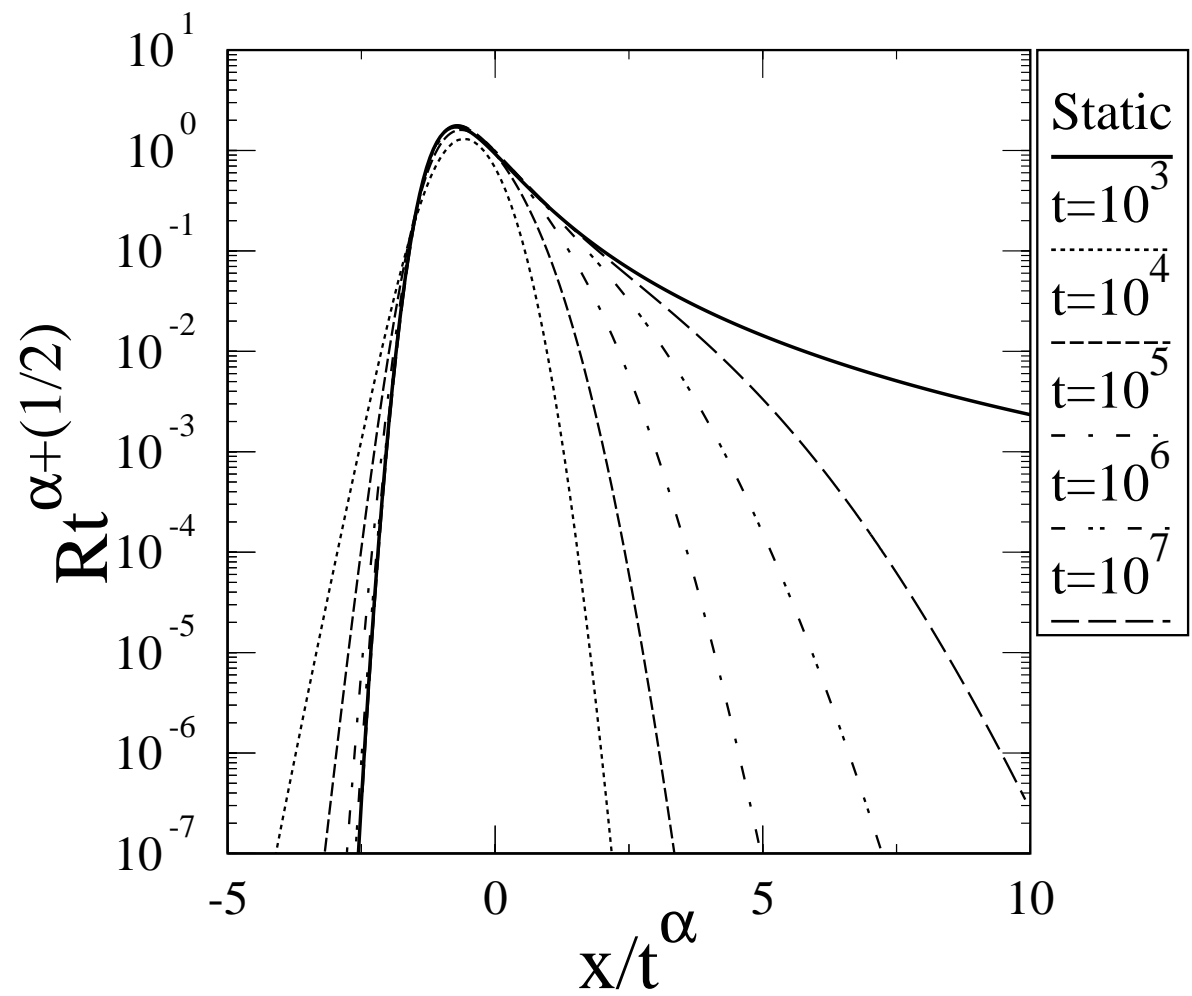
\begin{figure}
\caption{$R \cdot t^\beta$ vs.\ $x/t^\alpha$ for $n = 1$ and $m =4$
and $t = 10^3, 10^4,\ldots,10^7$.}
\end{figure}

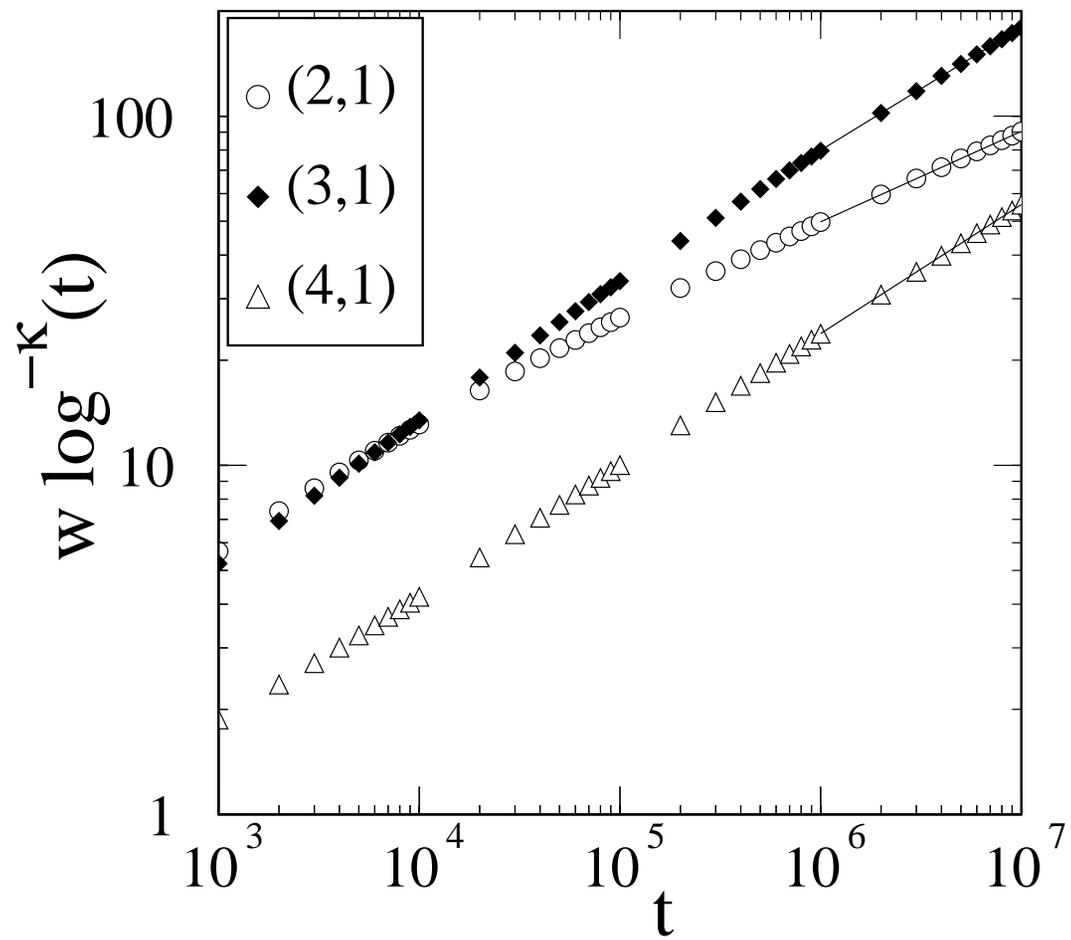
\begin{figure}
\caption{
Log-log plot of $w\log^{-\kappa}(t)$ versus $t$.
The straight lines are a least-squares fit to the last decade.}
\end{figure}

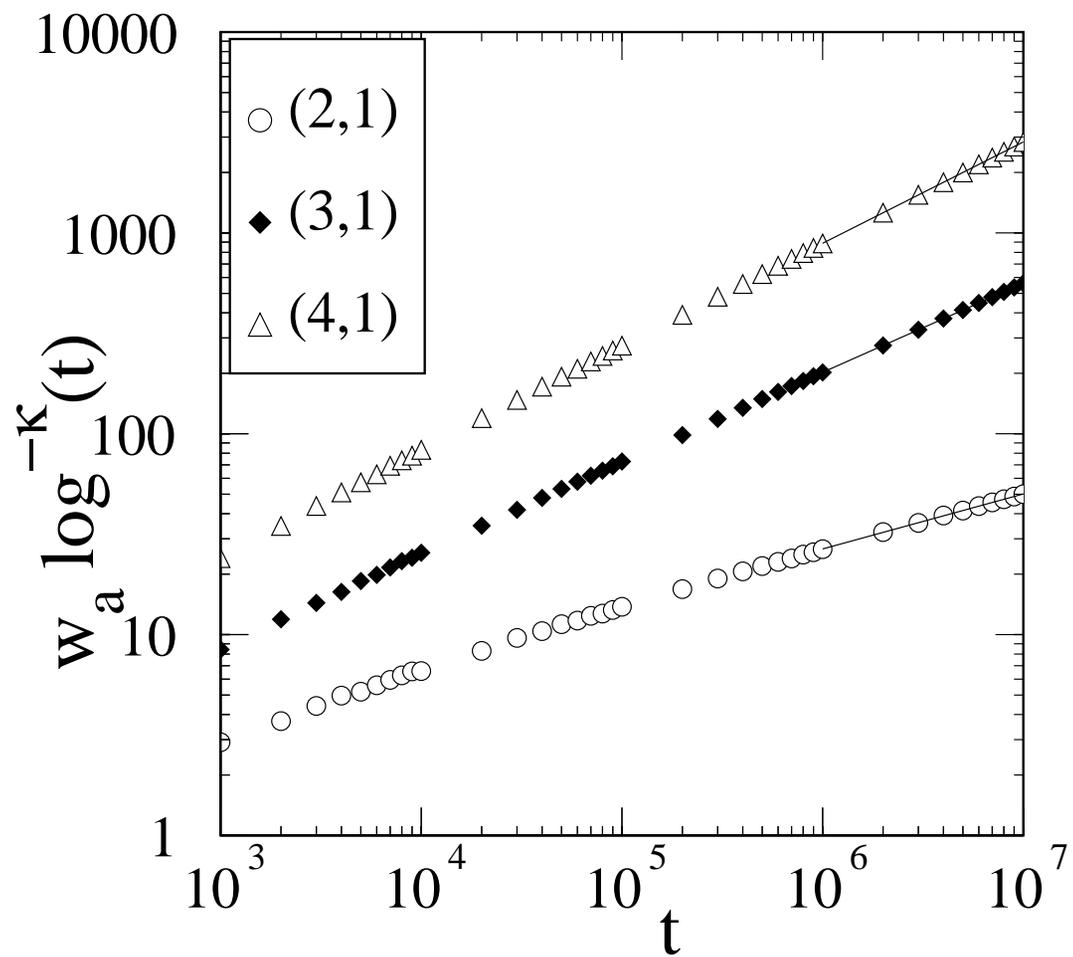
\begin{figure}
\caption{Log-log plot of $w_a\log^{-\kappa}(t)$ versus $t$.
The straight lines are a least-squares fit to the last decade.}
\end{figure}

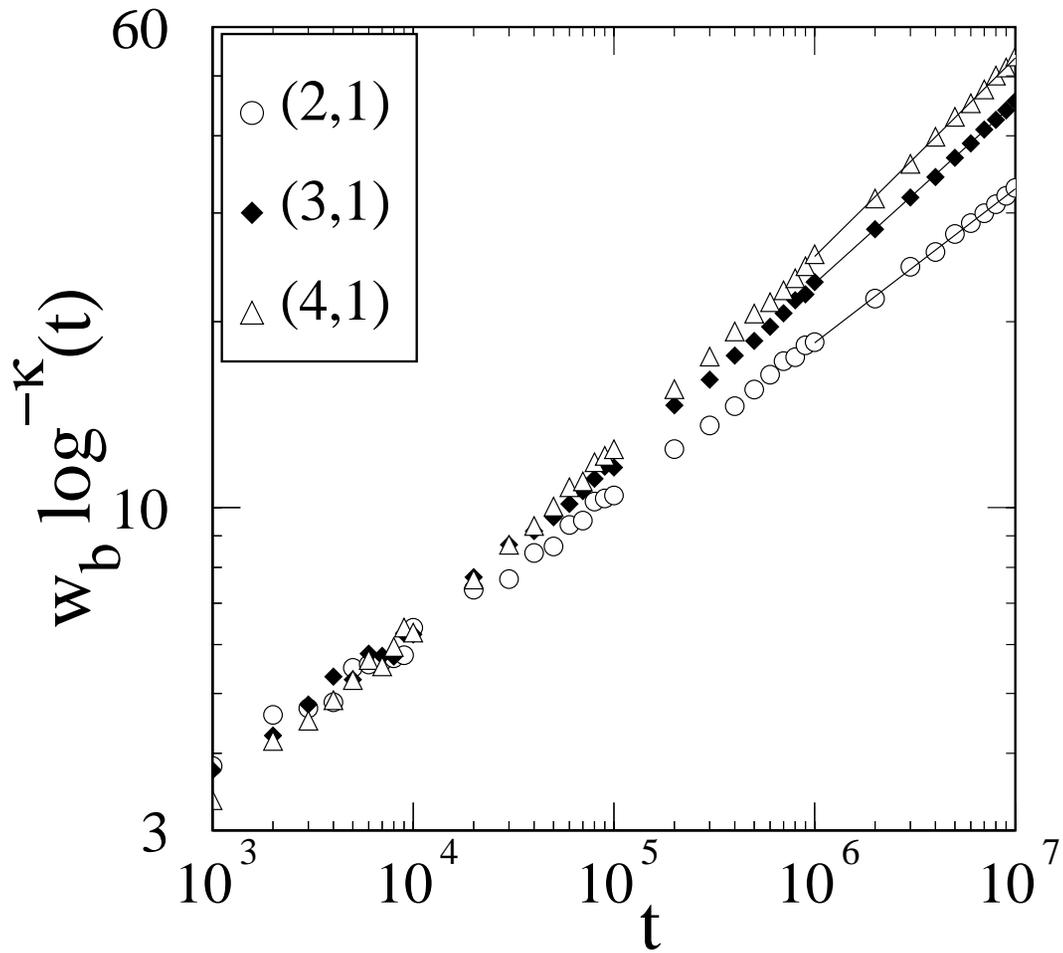
\begin{figure}
\caption{
Log-log plot of $w_b\log^{-\kappa}(t)$ versus $t$.
The straight lines are a least-squares fit to the last decade.}
\end{figure}

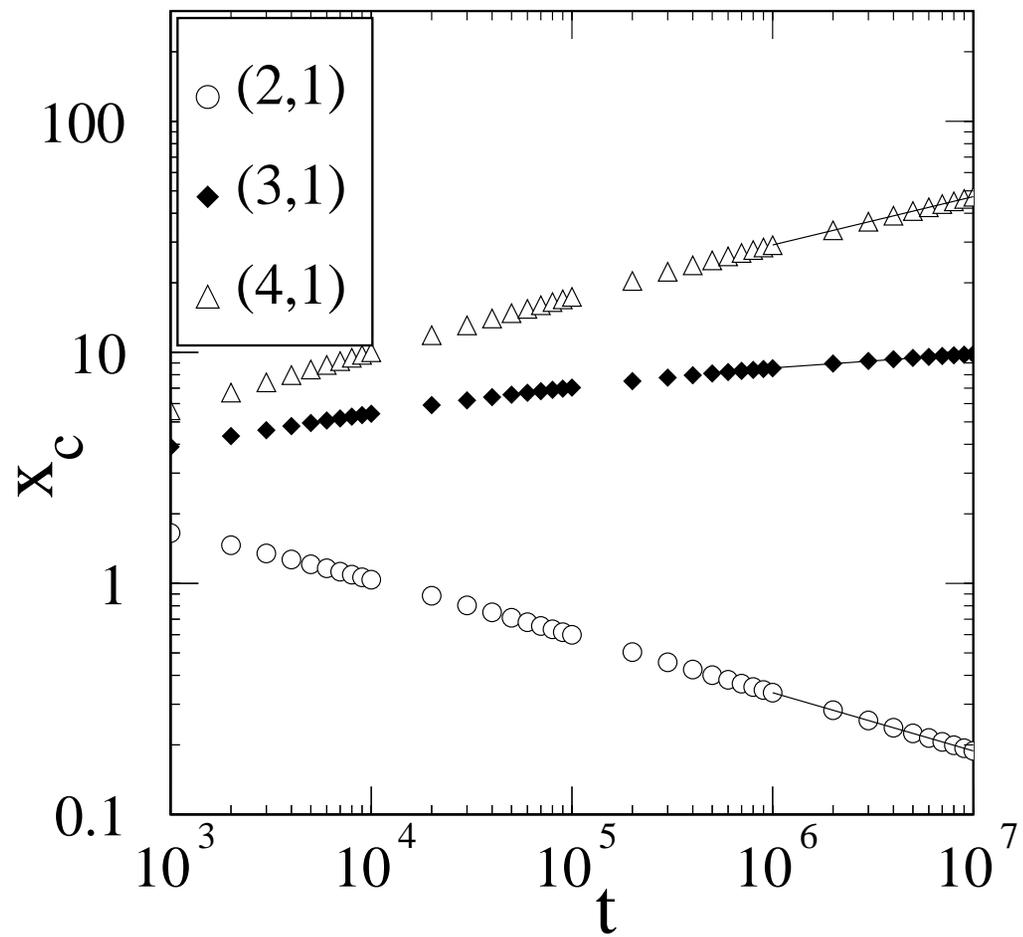
\begin{figure}
\caption{Log-log plot of $x_c\log^{-\kappa}(t)$ versus $t$.
The straight lines are a least-squares fit to the last decade.}
\end{figure}

\end{document}